# Tera-sample-per-second arbitrary waveform generation in the synthetic dimension

Yiran Guan[1], Jiejun Zhang[1]*, Lingzhi Li[1], Ruidong Cao[1], Guangying Wang[1], Jingxu Chen[1] & Jianping Yao[1,2]*

[1]Guangdong Provincial Key Laboratory of Optical Fiber Sensing and Communications, Institute of Photonics Technology, Jinan University, Guangzhou, 511443, China

[2]Microwave Photonics Research Laboratory, School of Electrical Engineering and Computer Science, University of Ottawa, Ottawa, ON K1N 6N5, Canada.

*Correspondence to: zhangjiejun@jnu.edu.cn, jpyao@uottawa.ca

## Abstract

The synthetic dimension opens new horizons in quantum physics and topological photonics by enabling new dimensions for field and particle manipulations. The most appealing property of the photonic synthetic dimension is its ability to emulate high-dimensional optical behavior in a unitary physical system. Here we show that the photonic synthetic dimension can transform technical problems in photonic systems between dimensionalities, providing unexpected solutions to technical problems that are otherwise challenging. Specifically, we propose and experimentally demonstrate a photonic Galton board (PGB) in the temporal synthetic dimension, in which the temporal high-speed challenge is converted into a spatial fiber-optic length matching problem, leading to the experimental generation of tera-sample-per-second arbitrary waveforms. Limited by the speed of the measurement equipment, waveforms with sampling rates of up to 341.53 GSa/s are recorded. Our proposed PGB operating in the temporal synthetic dimension breaks the speed limit in a physical system, bringing arbitrary waveform generation into the terahertz regime. The concept of dimension conversion offers possible solutions to various physical dimension-related problems, such as super-resolution imaging, high-resolution spectroscopy, time measurement, etc.

# 1. Introduction

The synthetic dimension provides additional dimensions for field and particle operations to explore higher dimensional physical phenomena in a lower-dimensional physical system[1,2]. Recently, synthetic dimension has opened a new perspective in photonic systems[3,4]. By dynamically modulating optical modes[5-7], orbital angular momenta[8,9], or by using multiple pulses[10-14], a photonic lattice in synthetic space can be constructed. In addition, for a photonic system having multiple optical parameter space, each parameter can be seen as an extra synthetic dimension[15,16]. Numerous photonic systems based on photonic synthetic dimension have been proposed for the implementation of optical quantum walks[17,18], band structure[5,14], topological photonics[10,19-21], and parity-time symmetry[16,22,23]. In 2020, Szameit et al. presented an efficient funnel for light in temporal synthetic space based on the non-Hermitian skin effect[10], in which the light field travels towards an interface regardless of its shape and input position. On the other hand, the Galton board invented by Galton in 1889 demonstrated the central limit theorem by observing the distribution of beads at the bottom of the board. It has been introduced to the optics community for studying spectral diffusion of a light wave[24] and quantum phenomenon in nonlinear optics[25-27]. The optical funnel is similar to a special Galton board, in which the drop position of the beads is controlled.

Microwave arbitrary waveforms with a wide bandwidth have been widely used in modern radar and microwave imaging systems to increase the range and imaging resolution[28-31]. Due to the limited sampling rate of an electronic analog-to-digital converter (ADC), a state-of-the-art electronic arbitrary waveform generator (AWG) using an electronic ADC can operate with a sampling rate up to 128 GSa/s and a maximum analog bandwidth of 65 GHz[32]. On the other hand, a photonic-assisted microwave arbitrary waveform generator can operate at a speed far beyond 128 GSa/s thanks to the inherent high speed and broad bandwidth offered by modern photonics. In an arbitrary waveform generation system based on optical pulse shaping, a spatial light modulator (SLM)[29,33,34], metasurface[35,36] or other spatial mask[37] is used to shape the optical spectrum of an ultra-short optical pulse, to tailor the magnitude and phase of the spectrum corresponding to the temporal waveform to be generated. However, those systems are implemented based on free-space optics, making the systems bulky, costly, and vulnerable to environmental disturbances. In recent years, benefiting from the rapid progress in microwave photonics[38-40], microwave arbitrary waveform generation with lower loss and smaller size based on optical pulse shaping or optical digital-to-analog converter[41-47] implemented by fiber-optic or integrated devices[48,49] in the temporal[50-52] or frequency domain[53-55] has been reported. Those AWGs have better waveform reconfigurability, but the bandwidth of a generated microwave waveform is still small, limited by the bandwidths of electro-optical modulators and photodetectors. Synthetic dimension is one potential technique to address this challenge, providing a potential solution to generate arbitrary waveforms at a much higher sampling rate and much wider bandwidth, and at the same time, better reconfigurability.

Here we show that a photonic Galton board (PGB) in the temporal synthetic dimension is proposed and experimentally demonstrated, in which a temporal high-speed challenge is converted into a spatial fiber-optic length matching problem, leading to the realization of tera-sample-per-second arbitrary waveform generation. Specifically, the PGB is implemented based on two coupled fiber loops with different time delays and one acousto-optic modulator (AOM) incorporated in each of the two loops, to control the gain or loss of the optical pulse recirculating in the loop, resulting in a synthesized one-dimensional (1D) temporal photonic system. The distribution of the beads at the bottom of the PGB corresponds to the shape of the generated waveform, which is defined by the megahertz modulation signals applied to the AOMs. Tera-sample-per-second arbitrary waveforms are experimentally generated when the duration of the seed pulse and the time delay difference of the two loops are both less than 1 ps. Limited by the speed of the measurement equipment, waveforms with sampling rates of up to 341.53 GSa/s are recorded. Our proposed PGB operating in the temporal synthetic dimension breaks the speed limit in a physical system, bringing arbitrary waveform generation into the terahertz regime. The concept of dimension conversion offers possible solutions to various physical dimension-related problems, such as super-resolution imaging, high-resolution spectroscopy, time measurement, etc.

## 2. Principle

A traditional Galton board has multiple rows of pegs equally separated by a given interval. When a bead reaches a peg, it will be bounced to the right or left with a fixed probability of 50%. At the bottom of the Galton board, a Gaussian distribution of the beads is produced. In a PGB, as shown in Figure 1, multiple pegs with each having a tunable coupling ratio are used to bounce a photon to the left or right. $T_1$ and $T_2$ are the time delays of the left and right paths, respectively. At the bottom of the PGB, the temporal distribution of the photons is produced which is a temporally sampled arbitrary waveform with its shape and sampling rate determined by the coupling ratio and path length difference: $T_2-T_1$, respectively, so the challenge of a high sampling rate is subtly converted into a fiber-optic length matching problem.

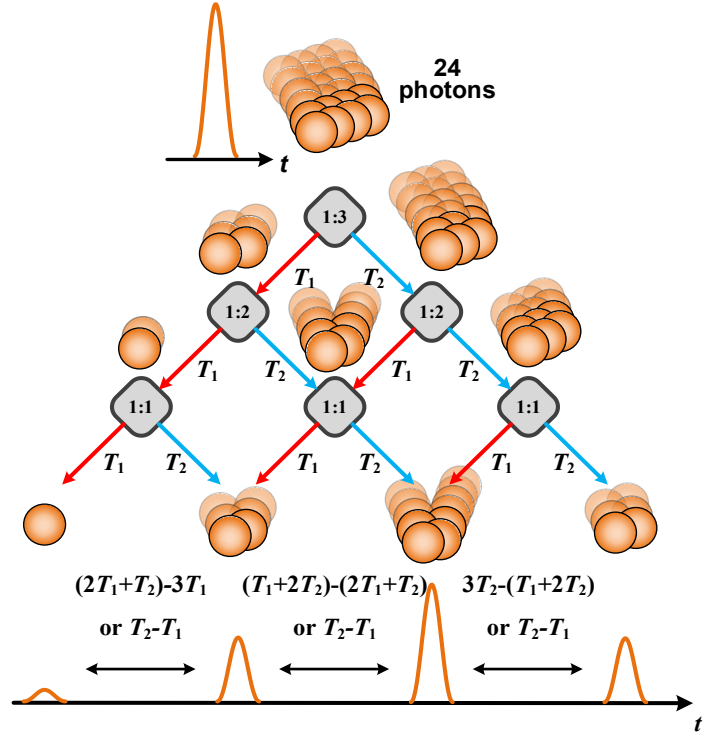

**Fig.1. Schematic diagram of the proposed PGB.** The pegs (gray rounded rectangles) of each row have a tunable coupling ratio (the number in the gray rounded rectangles, corresponding to the probability), thus they can be made to control the number of photons (beads, orange circles) going to the left or right. $T_1$ and $T_2$ are the time delays of the left and right paths, respectively. At the bottom of the PGB, the temporal distribution of the photons is formed, which is a temporally sampled arbitrary waveform whose shape and sampling interval are determined by the pegs with tunable coupling ratio and path length difference, respectively.

Figure 2(a) shows a dual-loop fiber-optic system consisting of two coupled loops having different loop lengths with an AOM incorporated in each loop for the implementation of the PGB in the temporal synthetic dimension. An ultra-short pulse containing multiple photons (beads) from a mode-locked laser (MLL) with a temporal width $d$ is inputted into the long loop (blue line) through an optical coupler (OC). The photons are split into two parts by a beam splitter (BS). The BS has a fixed splitting ratio, but the joint operation of the BS and the AOMs corresponds to a tunable BS with the splitting ratio controlled by the gain and loss of the AOMs. Then, the two groups of photons, i.e., two pulses, return to the BS and are split into four parts corresponding to four pulses. As the seed pulse recirculates in the two loops of different lengths for more round trips, at the output of each loop, a sequence of pulses is generated. By detecting the optical pulse sequences from the two loops at two photodetectors (PDs) and combining the detected electrical pulse sequences, the temporal distribution of the photons is produced, which is a temporally sampled arbitrary waveform with its shape determined by the tunable coupling ratio during different round trips. The time interval between adjacent pulses in the generated pulse sequence is the time delay difference between the two loops, so the challenge of a high sampling rate is subtly converted into a fiber-optic length matching problem. The equivalence of the fiber optic system in Fig. 2(a) to a PGB is shown in Fig. 2(b). The *n*-th pulse in the *m*-th round trip after being modulated by the AOM with an electric field $u_{m,n}$ in the long loop and $v_{m,n}$ in the short loop is coupled to its nearest neighbor sites $u(v)_{m+1,n+1}$ and $v(u)_{m+1,n}$ in the next round trip, respectively. Thus, a synthetic temporal dimension, pulse position $n$, is synthesized. We modulate the seed pulse in the physical dimension, round trip $m$, at kilohertz or megahertz, an ultra-high-speed waveform of terahertz can be generated along the synthetic dimension, thus we successfully transform technical problems in the photonic system between dimensionalities.

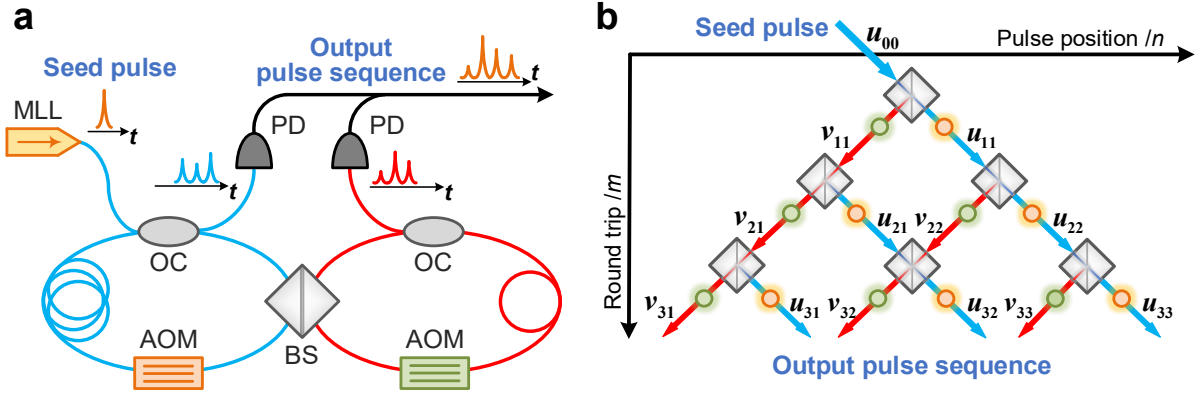

**Fig. 2. A dual-loop fiber-optic system for the implementation of the proposed PGB in the temporal synthetic dimension.** (a) An ultra-short pulse containing multiple photons from a mode-locked laser (MLL) is inputted into the long loop (blue line) through an optical coupler (OC). The long loop and the short one (red line) are mutually coupled by a beam splitter (BS). Each loop contains an acousto-optic modulator (AOM) for controlling the gain and loss of the pulses recirculating in the loops. After the seed pulse recirculates in the two loops for more round trips, a sequence of pulses is generated from each loop. By detecting the optical pulse sequences and combining them, a temporally sampled arbitrary waveform is generated. (b) The equivalence of the fiber optic system in (a) to the proposed PGB. The blue and red arrowed lines indicate the pulses traveling in the long and short loops, respectively. The BS is depicted as the gray rectangle and the AOMs in the long and short loop are depicted as orange and green circles, respectively. The $n$-th pulse in the $m$-th round trip with an electric field $u_{m,n}$ in the long loop and $v_{m,n}$ in the short loop is coupled to its nearest neighbor sites $u(v)_{m+1,n+1}$ and $v(u)_{m+1,n}$ in the next round trip, respectively. Thus, a synthetic temporal dimension, pulse position $n$, is synthesized. PD: photodetector.

The electric field of the optical pulse at the output of the two loops can be expressed by

$$u_{m+1,n} = G_{u,m+1}\left(\frac{\sqrt{2}}{2}u_{m,n-1} + i\frac{\sqrt{2}}{2}v_{m,n}\right) \quad (1.1)$$

$$v_{m+1,n} = G_{v,m+1}\left(i\frac{\sqrt{2}}{2}u_{m,n-1} + \frac{\sqrt{2}}{2}v_{m,n}\right) \quad (1.2)$$

where $G_{u,m+1}$ and $G_{v,m+1}$ are the gain/loss of the AOMs in the long and short loop for the $(m+1)$-th round trip, respectively. Thus, as the seed pulse recirculates in the two loops of different lengths for more round trips, a temporally sampled arbitrary waveform $s[n]$ is generated at the bottom of the PGB, which can be expressed by

$$s[n] = |u_{m,n}|^2 + |v_{m,n+1}|^2, \quad 0 \leq n \leq m \quad (2)$$

where $u_{m,0}=0$ and $v_{m,m+1}=0$ (see Supplementary Material 1). The sampling interval is equal to the time delay difference between the two loops which can be made ultra-small. The smallest sampling interval is only limited by the temporal width of the recirculating pulse at the *m*-th round trip, which is broadened by the chromatic dispersion in the loops and given by

$$d' = D \times \Delta\lambda \times m \times l \tag{3}$$

where $D$ is the dispersion coefficient of the loops, $\Delta\lambda$ is the spectrum width of the seed pulse and $l$ is the length of the long loop. Thus, the sampling rate $f_s$ of the generated temporally sampled waveform can be expressed by

$$f_s = \frac{1}{T_2 - T_1} = \frac{1}{\Delta t} \leq \frac{1}{d'} \tag{4}$$

where $d'$ is the temporal width of the recirculating pulse. Eq. (4) indicates that the sampling rate of the generated waveform from the PGB can be ultra-high when the time delay difference is ultra-small and the maximum sampling rate is only limited by the temporal width of the recirculating pulse.

## 3. Numerical and experimental results

A numerical study is performed to evaluate the operation of the proposed PGB for the generation of arbitrary waveforms. We set a constant loss, $G_{v,m}=0.5$, in the short loop, and a variable loss, $G_{u,m} \in (0,1]$, in the long loop for different round trips. We develop an *Auto-Fit* system with a backpropagation algorithm used to calculate $G_{u,m}$ for the target waveform (see Supplementary Material 2). The generation of six waveforms with 11 sampling points and two waveforms with 31 sampling points is presented. Figure 3(a1)-(h1) shows the eight target waveforms. By calculating $G_{u,m}$ of the target waveforms and applying them to the AOM in the long loop, the desired waveforms are generated, which are shown in Fig. 3(a2)-(h2). The sampling interval between two adjacent sampling points equals the time delay difference $\Delta t$ between the two loops. When both the temporal duration of every sampling point and the time delay difference $\Delta t$ are less than 1 ps, the sampling rate of the generated arbitrary waveform can reach terahertz.

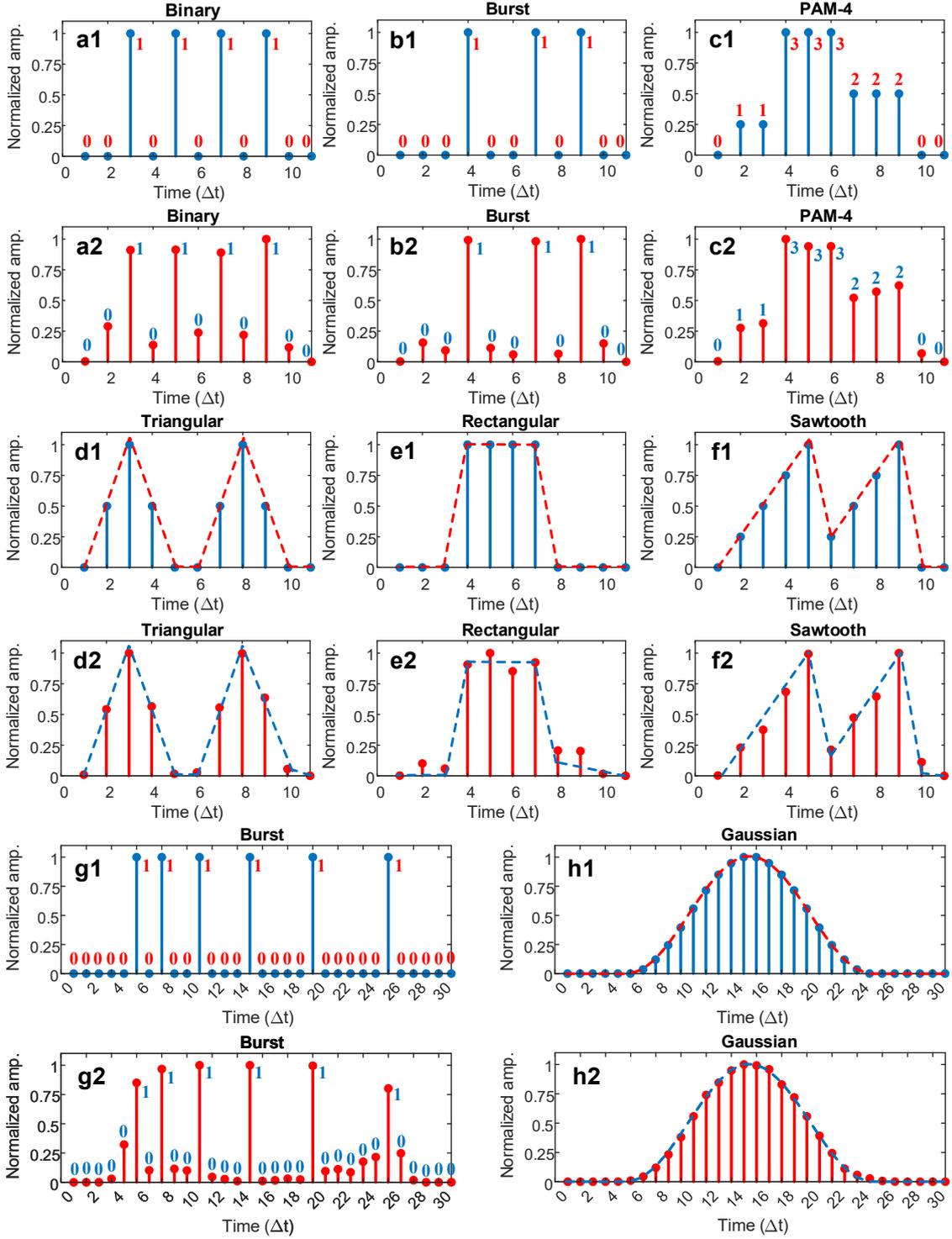

**Fig. 3. Target waveforms and the corresponding simulation results.** The target waveforms with 11 sampling points: (a1) binary coded, (b1) burst, (c1) PAM-4, and their simulated results in (a2), (b2), and (c2), respectively. The target waveforms with 11 sampling points: (d1) triangular, (e1) rectangular, and (f1) sawtooth waveforms and their simulated results in (d2), (e2), and (f2), respectively. The target waveforms with 31 sampling points: (g1) burst and (h1) Gaussian waveforms and their simulated results in (g2) and (h2), respectively. Note that the generated waveforms are sampled and can be a continuous-time analog waveform after passing through a lowpass filter.

For our experimental investigation of our proposed PGB, we develop an all-fiber-based dual-loop system, as shown in Figure 4. An optical pulse train is generated by the MLL. The seed pulse is cut off from this pulse train through a Mach-Zehnder modulator (MZM) and broadened to have a 3-dB temporal width of 31 ps or a 10 dB width of 56 ps through the SMF which has a length of 330 m (see Supplementary Material 3). Then, it is injected into the dual-loop system, where the short loop has a length of 34.6 m, and a tunable delay line (TDL) is incorporated in the long loop, which is used to tune the time delay difference between the two loops and hence the sampling rate of the generated waveforms. In each loop, an erbium-doped fiber amplifier (EDFA) and a tunable optical filter (TOF) are used to ensure the seed pulse recirculates in the system for a sufficient number of round trips and to maintain a high signal-to-noise-ratio (SNR). The polarization controllers (PCs) are used to minimize polarization-dependent loss. Finally, an optical coupler (OC) is used in each loop to direct the pulse sequence to a detection unit (DU), which consists of an AOM employed to apply a designated time window, a photodetector (PD) for optical-to-electrical conversion, and a high-speed oscilloscope (OSC). A temporally sampled arbitrary waveform is generated by combining the detected electrical pulse sequences from the two DUs.

In the short loop, we keep $G_{v,m} = 0.5$ by fixing the driving signal applied to the AOM to be a sinusoidal signal at 200 MHz with a fixed amplitude. At the same time, a second sinusoidal signal with a carrier frequency of 80 MHz amplitude modulated by $G_{u,m}$ at 5.78 MHz is applied to the AOM as a driving signal in the long loop (see Supplementary Material 3). After the seed pulse recirculating in the PGB for 10 round trips, waveforms with 11 sampling points are generated, including a binary coded, a burst, a PAM-4, a triangular, a rectangular, and a sawtooth waveform, as shown in Fig. 5(a) to (f). The time delay difference between the two loops is 149.93 ps, corresponding to a sampling rate of 6.67 GSa/s. A burst and a Gaussian waveform with 31 sampling points are also generated after the $30^{th}$ round trip, as shown in Fig. 5(g) and (h), respectively. In these cases, the time delay difference between the two loops is 168.07 ps, which results in a sampling rate of 5.95 GSa/s. We calculate the root mean square error (RMSE) of the generated waveforms to evaluate the performance of the PGB for the generation of waveforms, which shows an average RMSE of 0.1820 (see Supplementary Material 4), confirming the good fidelity of the generated waveforms.

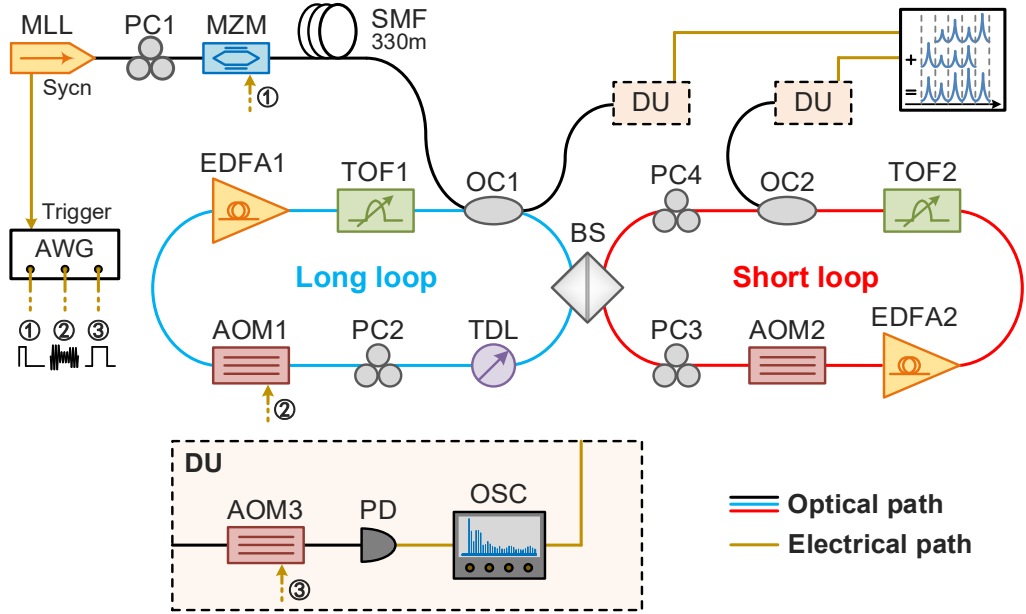

**Fig. 4. Experimental setup of the PGB in the temporal synthetic dimension.** The long loop (blue line) and the short loop (red line) are mutually coupled by a beam splitter (BS). A tunable delay line (TDL) is incorporated in the long loop to control the time delay difference and hence the sampling rate of the generated waveform. The temporally sampled arbitrary waveform is generated by combining the detected electrical pulse sequences after the two detection units (DUs). Three signals generated from an AWG are applied to the MZM, AOM1, and AOM3 for cutting off the pulse train, modulating the recirculating pulse sequence, and extracting the pulse sequence, respectively. PC: polarization controller; MZM: Mach-Zehnder modulator; SMF: single mode fiber; OC: optical coupler; AOM: acousto-optic modulator; EDFA: Erbium-doped fiber amplifier; TOF: tunable optical filter; PD: photodetector; OSC: oscilloscope.

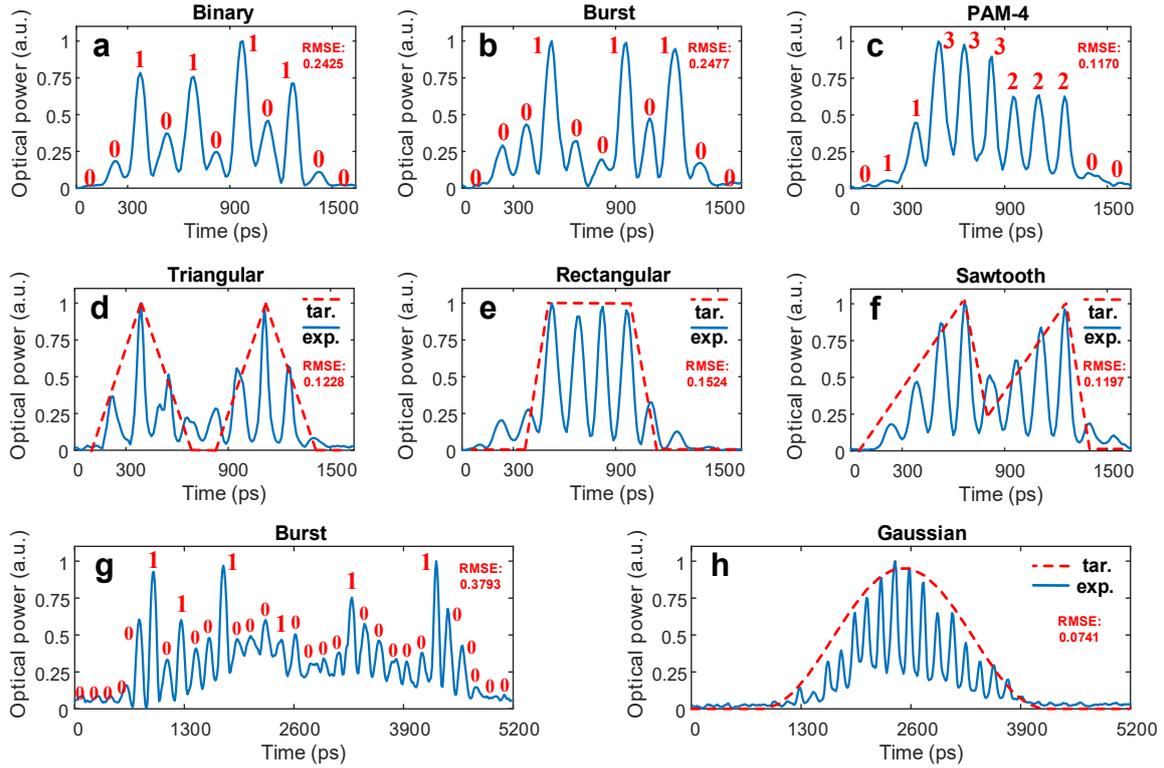

**Fig. 5. Experimentally generated arbitrary waveforms at 6.67 and 5.95 GSa/s.** Generated waveforms (solid blue line) with 11 sampling points: (a) binary coded, (b) burst, (c) PAM-4, (d) triangular, (e) rectangular, and (f) sawtooth waveforms. The time delay difference between the two loops is 149.93 ps, corresponding to a sampling rate of 6.67 GSa/s. Generated waveforms with 31 sampling points: (g) burst and (h) Gaussian waveforms. The time delay difference is 168.07 ps, corresponding to a sampling rate of 5.95 GSa/s. The target waveforms are also shown in (d), (e), (f) and (h) as dashed red lines for comparison. The RMSE shows an average value of 0.1820.

In order to further explore the capability of the system in performing faster-speed waveform generation, a second experiment is performed. In this case, a pulse after the MZM is directly injected into the long loop without passing through the SMF. To eliminate the chromatic dispersion in the two loops, a dispersion compensating fiber (DCF) of about 7 m is employed in each of the two loops. Then, we tune the time delay difference between the two loops from 46.08 to 12.50 ps, corresponding to a tunable sampling rate from 21.7 to 80.0 GSa/s. Four analog waveforms are generated and shown in Figure 6, in which the dashed red lines and solid blue lines are the target waveforms and the experimentally generated waveforms, respectively. The corresponding RMSE is calculated to have an average value of 0.0798 (see Supplementary Material 4), again confirming the good fidelity of the generated waveforms.

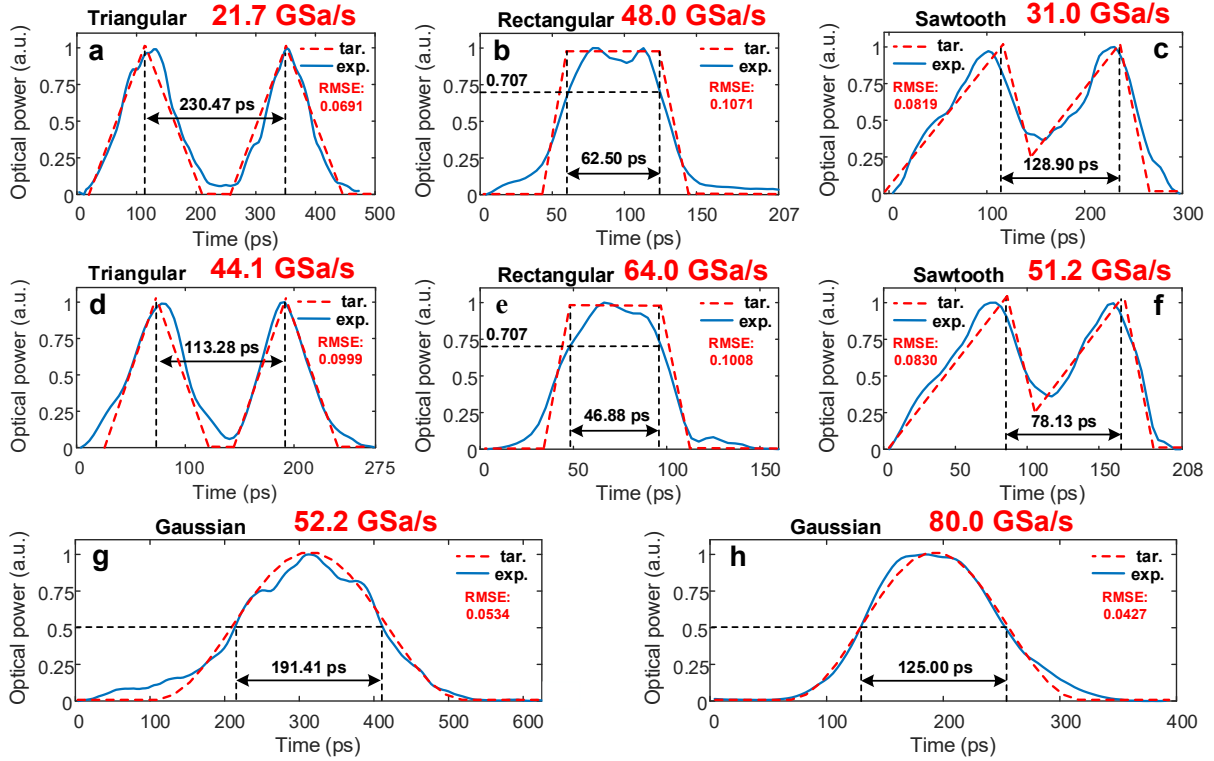

**Fig. 6. Experimentally generated arbitrary waveforms with high sampling rates ranging from 21.7 to 80.0 GSa/s.** The generated waveforms (solid blue lines) with 11 sampling points at different sampling rates: (a) triangular at 21.7 GSa/s, (b) rectangular at 48.0 GSa/s, sawtooth waveforms at 31.0 GSa/s, (d) triangular at 44.1 GSa/s, (e) rectangular at 64.0 GSa/s, sawtooth waveforms at 51.2 GSa/s. The generated Gaussian waveforms with 31 sampling points: (g) Gaussian at 52.2 GSa/s and (h) Gaussian at 80.0 GSa/s. The target waveforms are also shown as dashed red lines for comparison. The RMSE shows an average value of 0.0798.

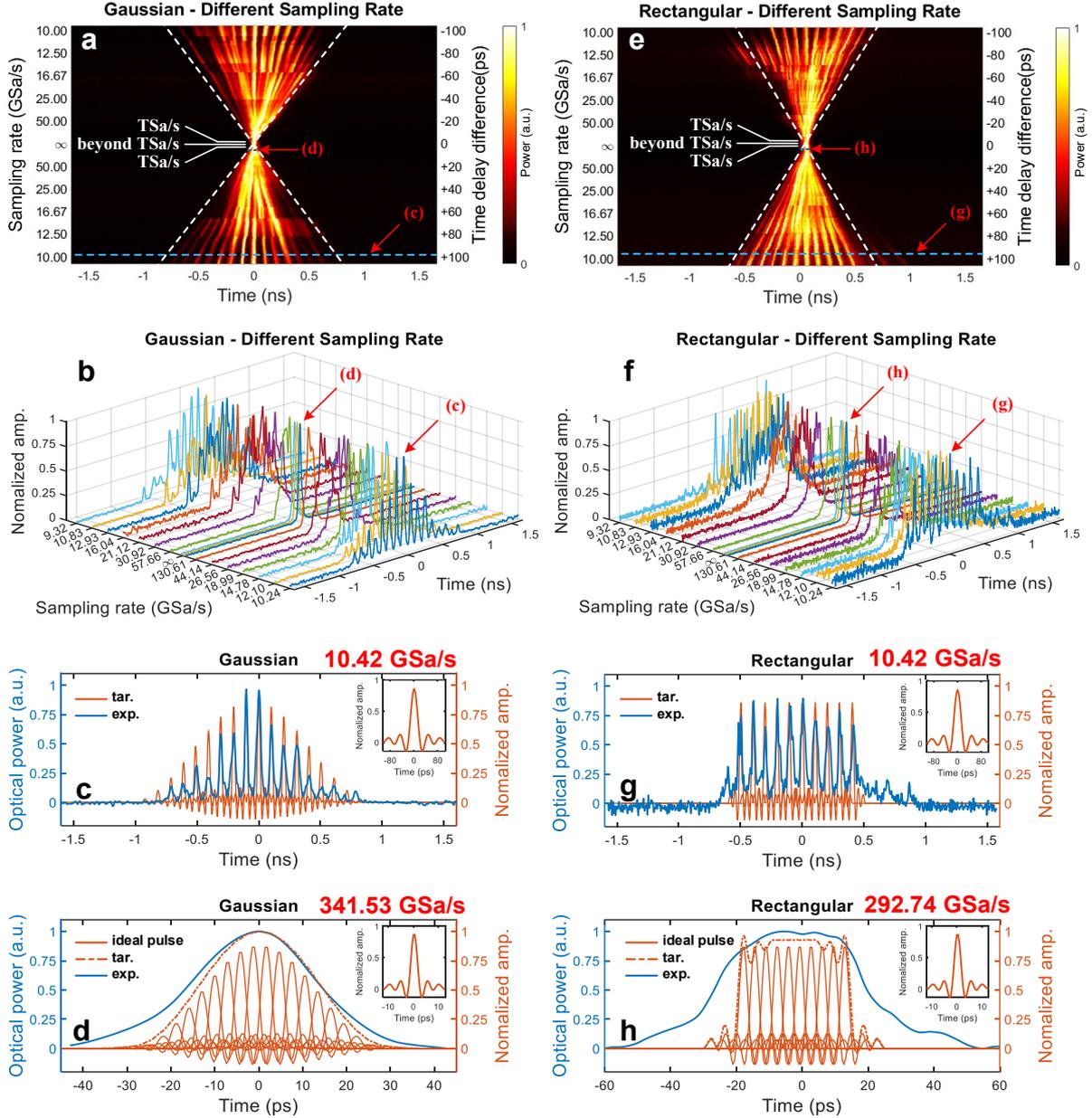

**Fig. 7. Sampling rate tuning.** (a) The evolution of an experimentally generated Gaussian waveform when the time delay difference between the two loops is tuned from +100 to -100 ps. The dashed white lines show the evolution of the waveform width to indicate the variation of the sampling rate. At the area near the intersection of the two dashed white lines, the sampling rate can reach a terahertz or higher when the time delay difference is within +1 and -1 ps. (b) The evolution of fifteen generated Gaussian waveforms in (a) is shown in the 3D plot. (c) A generated Gaussian waveform in solid blue lines at 10.42 GSa/s. A target waveform in solid orange lines is also shown for comparison. A zoom-in view of an individual pulse in the target waveform is given in the inset by which the 3 dB temporal width is measured to be 23.8 ps. (d) A generated Gaussian waveform in a solid blue line at 341.56 GSa/s. The dashed orange line shows the target waveform, which is generated by an ideal pulse burst shown in solid orange lines by passing the pulse burst through a lowpass filter. The insert is a zoom-in view of an individual pulse with a 3 dB temporal width of 2.86 ps in the ideal pulse burst. (e)-(h) show the generation of a rectangular waveform at different sampling rates.

Then, the evolution of experimentally generated waveforms when the time delay difference is tuned is studied. Figure 7(a) shows the evolution of an experimentally generated Gaussian waveform when the time delay difference is tuned from +100 to -100 ps. Note the long loop will become a short loop if the time delay difference becomes negative. The sampling rate is increased from 10.00 GSa/s to infinity and then decreased from infinity back to 10.00 GSa/s. The dashed white lines in Fig 7(a) show the evolution of an experimentally generated Gaussian waveform when the time delay difference between the two loops is tuned from +100 to -100 ps. The dashed white lines show the evolution of the temporal width to indicate the variation of the sampling rate. As can be seen, if the time delay difference is reduced to 1 ps or less, the sampling rate can reach a terahertz or higher. Fig. 7(b) shows the evolution of the generated Gaussian waveforms in Fig. 7(a) in a 3D plot, where fifteen Gaussian waveforms with different sampling rates are shown. Fig. 7(c) shows a generated Gaussian waveform in solid blue lines at 10.42 GSa/s. A target waveform in solid orange lines is also shown for comparison. A zoom-in view of an individual pulse in the target waveform is given in the inset by which the 3 dB temporal width is measured to be 23.8 ps. Fig. 7(d) shows a generated Gaussian waveform in a solid blue line at 341.56 GSa/s. The dashed orange line shows the target waveform, which is generated by passing an ideal pulse burst shown in solid orange lines through a lowpass filter, which is a PD (due to its finite bandwidth) in the experiment. The insert is a zoom-in view of an individual pulse with a 3 dB temporal width of 2.86 ps in the ideal pulse burst. The generation of a rectangular waveform with 31 sampling points at different sampling rates is also performed and the results are shown in Fig 7(e)-(h). These experimental results are all achieved by using a PD with a bandwidth of 50 GHz and a high-speed OSC with a bandwidth of 80 GHz, a sampling rate of 256 GSa/s, and a minimum rise/fall time of 5.6 ps (10-90%) and 3.9 ps (20-80%).

## 4. Discussion and conclusion

We showed that the photonic synthetic dimension can transform technical problems in photonic systems between dimensionalities, providing unexpected solutions to technical problems that are otherwise challenging. Our proposed PGB operating in the temporal synthetic dimension breaks the speed limit in a physical system, bringing arbitrary waveform generation into the terahertz regime. The concept of dimension conversion offers possible solutions to various physical dimension-related problems, such as time measurement, high-resolution spectroscopy, super-resolution imaging, etc. By converting the temporal interval, spectral component, and space distance problem into the other dimensions or axes in a physical system, many ultimate technical issues may be subtly solved and profoundly pushed to new horizons.

In the proposed PGB, the sampling rate of the generated arbitrary waveform is determined by the time delay difference and is limited by the temporal width of the seed pulse and the dispersion in the two loops. In the experiment, the pulse recirculating in the two loops is cumulatively broadened by residual dispersion, which cannot be fully compensated, making the sampling rate limited. In addition, the bandwidths of the PDs and the OSC will also limit the maximum sampling rate. For a waveform with a sampling rate beyond 1 TSa/s, the bandwidths of the PDs and the OSC need to be over 500 GHz. In our experiment, the bandwidths of the PD and the OSC are 50 and 80 GHz, respectively, limiting the correctly recordable waveforms to a few hundreds of gigahertz, although the PGB has generated tera-sample-per-second waveforms.

In conclusion, we have proposed and experimentally demonstrated a PGB in the temporal synthetic dimension, in which a temporal high-speed challenge is converted into a spatial fiber-optic length matching problem, leading to the generation of tera-sample-per-second arbitrary waveforms. The PGB was implemented in an all-fiber-based dual-loop system, where two coupled fiber loops with different time delays and one AOM were incorporated into each of the two loops, to control the gain or loss of the optical pulse recirculating in the loop. The distribution of the beads at the bottom of the PGB corresponded to the shape of the generated waveform, which was defined by the megahertz modulation signals applied to the AOMs. By reducing the time delay difference between the two loops, the sampling rate can be increased. On the other hand, the maximum sampling rate is also limited by the temporal width of the recirculating optical pulse, which is affected by the temporal width of the seed pulse and the dispersion in the loops. Thus, when the seed pulse duration and the time delay difference of the two loops are less than 1 ps, tera-sample-per-second arbitrary waveforms were generated. Compared with the generation of arbitrary waveforms by pulse shaping in free-space optics[29,33-37], our proposed PGB in the synthetic dimension was implemented in fiber optics, so that it can have a smaller size and lower loss. Our proposed PGB breaks the bandwidth limitation of optoelectronic components such as electro-optical modulators, waveshapers, etc.[30,50-52,54]. It has the potential to be integrated based on silicon or InP platform with further reduced size and better stability for commercial applications[39,48,49,53].

## 5. Methods

**Implementation of the PGB in the synthetic dimension**
The PGB in the temporal synthetic dimension for the generation of ultra-high speed arbitrary waveforms is implemented using commercial off-the-shelf optical and optoelectronic components. An optical pulse train with a repetition rate of 20 MHz is generated by a femtosecond laser source (CALMAR OPTCOM FPL-03CFFJNU). Before being injected into the dual-loop fiber optic system, the repetition rate of the pulse train is reduced to 50 kHz through an MZM (FUJITSU H74M-5208-J048) to make two adjacent pulses in the pulse train have sufficiently large time spacing to allow a pulse to recirculate in the loops for multiple times without overlapping with a second pulse from the pulse train. The TDL (General Photonics MDL-002) with a maximum time delay of 560 ps is used for controlling the time delay difference between the two loops so that the sampling rate of the generated arbitrary waveform can be tuned. The AOM1 (CETC SGTF80-1550-1) with a shifting frequency of 80 MHz, the EDFA1 (Max-Ray Photonics EDFA-BA-20-B) bumped at about 75 mA, and the TOF1 (santec OTF-350) with a center wavelength of 1553.11 nm and a passband of 3.94 nm are incorporated in the long loop. In the short loop, the AOM2 (Brimrose TEM-210-50-10-1550-2FP) has a shifting frequency of 200 MHz, the EDFA2 (PYOE-EDFA-C) is bumped at about 66 mA and the TOF2 (Alnair Labs BVF-300CL) is also working at a center wavelength of 1553.11 nm and a passband of 3.94 nm. The EDFAs and the TOFs are used to compensate for the round-trip loss and to filter out the amplified spontaneous (ASE) noise from the EDFA outside the signal band. We adjust the PC, EDFA, and TOF in each loop to ensure the pulses recirculate in each loop for enough round trips. AOM3 in the DU acts as an optical switch for cutting off the optical pulse sequence at the $10^{th}$ or the $30^{th}$ round trip. The square signals applied to the MZM and the AOM3 in the DUs are generated by a two-channel AWG (RIGOL DG822).

**Generation of arbitrary waveforms**

The modulation signal applied to the AOM1 in the long loop is generated by another AWG (Tektronix AWG70002A). The output optical pulse sequence from each loop is converted into an electrical signal by the PDs ($u^2t$ XPDV21x0RA), which have a bandwidth of 50 GHz. A high-speed oscilloscope (Teledyne LabMaster 10-36Zi) with a bandwidth of 36 GHz and a sampling rate of 80 GSa/s is used to monitor the low-speed generated waveforms in Fig. 5. Another high-speed oscilloscope (Keysight UXR 0804A) with a bandwidth of 80 GHz and a sampling rate of 256 GSa/s is used in the case of monitoring faster-speed waveforms in Fig. 6 and Fig. 7.

The Sync port of the MLL is connected to the Trigger port of the two-channel arbitrary waveform generator (RIGOL DG822) for time synchronization. At the same time, the Sync port of this AWG is connected to another one (Tektronix AWG70002A).

# 6. Author contributions

Y. G. conceived the idea, designed the theoretical simulation, conducted the experiment, analyzed the data, and wrote the paper; J. Z. and J. Y. conceived the idea, directed the theoretical simulation, guided the experiment, and wrote the paper; L. L., R. C., G. W., and J. C. contributed to the experiment.

# 7. Acknowledgements

This work was supported by the National Natural Science Foundation of China (61905095, 61860206002, 61805103) and the Guangdong Province Key Field R&D Program Project (2020B0101110002).

# 8. References

1 Mancini, M. *et al.* Observation of chiral edge states with neutral fermions in

|   |   |
|---|---|
|   | synthetic Hall ribbons. *Science* **349**, 1510-1513, (2015). |
| 2 | Boada, O., Celi, A., Latorre, J. I. & Lewenstein, M. Quantum Simulation of an Extra Dimension. *Phys. Rev. Lett.* **108**, 133001, (2012). |
| 3 | Yuan, L., Lin, Q., Xiao, M. & Fan, S. Synthetic dimension in photonics. *Optica* **5**, 1396-1405, (2018). |
| 4 | Lustig, E. *et al.* Photonic topological insulator in synthetic dimensions. *Nature* **567**, 356-360, (2019). |
| 5 | Li, G. *et al.* Dynamic band structure measurement in the synthetic space. *Sci. Adv.* **7**, eabe4335, (2021). |
| 6 | Qin, C. *et al.* Spectrum control through discrete frequency diffraction in the presence of photonic gauge potentials. *Phys. Rev. Lett.* **120**, 133901, (2018). |
| 7 | Buddhiraju, S., Dutt, A., Minkov, M., Williamson, I. A. & Fan, S. Arbitrary linear transformations for photons in the frequency synthetic dimension. *Nat. Commun.* **12**, 2401, (2021). |
| 8 | Ouyang, X. *et al.* Synthetic helical dichroism for six-dimensional optical orbital angular momentum multiplexing. *Nat. Photon.* **15**, 901-907, (2021). |
| 9 | Luo, X., Zhang, C., Guo, G. & Zhou, Z. Topological photonic orbital-angular-momentum switch. *Phy. Rev. A* **97**, 043841, (2018). |
| 10 | Weidemann, S. *et al.* Topological funneling of light. *Science* **368**, 311-314, (2020). |
| 11 | Wimmer, M., Monika, M., Carusotto, I., Peschel, U. & Price, H. M. Superfluidity of light and its breakdown in optical mesh lattices. *Phys. Rev. Lett.* **127**, 163901, (2021). |
| 12 | Bartlett, B., Dutt, A. & Fan, S. Deterministic photonic quantum computation in a synthetic time dimension. *Optica* **8**, 1515-1523, (2021). |
| 13 | Schreiber, A. *et al.* A 2D Quantum Walk Simulation of Two-Particle Dynamics. *Science* **336**, 55-58, (2012). |
| 14 | Wimmer, M., Price, H. M., Carusotto, I. & Peschel, U. Experimental measurement of the Berry curvature from anomalous transport. *Nat. Phys.* **13**, 545-550, (2017). |
| 15 | Li, L. *et al.* Polarimetric parity-time symmetry in a photonic system. *Light Sci. Appl.* **9**, 169, (2020). |
| 16 | Zhang, J. *et al.* Parity-time symmetry in wavelength space within a single spatial resonator. *Nat. Commun.* **11**, 3217, (2020). |
| 17 | Chalabi, H. *et al.* Synthetic Gauge Field for Two-Dimensional Time-Multiplexed Quantum Random Walks. *Phys. Rev. Lett.* **123**, 150503, (2019). |
| 18 | Xiao, L. *et al.* Non-Hermitian bulk–boundary correspondence in quantum dynamics. *Nat. Phys.* **16**, 761-766, (2020). |
| 19 | Ozawa, T. & Price, H. M. Topological quantum matter in synthetic dimensions. *Nat. Rev. Phys.* **1**, 349-357, (2019). |
| 20 | Lustig, E. & Segev, M. Topological photonics in synthetic dimensions. *Adv. Opt. Photonics* **13**, 426-461, (2021). |
| 21 | Leefmans, C. *et al.* Topological dissipation in a time-multiplexed photonic resonator network. *Nat. Phys.* **18**, 442-449, (2022). |

22   Regensburger, A. *et al.* Parity-time synthetic photonic lattices. *Nature* **488**, 167-171, (2012).
23   Wimmer, M. *et al.* Observation of optical solitons in PT-symmetric lattices. *Nat. Commun.* **6**, 7782, (2015).
24   Bouwmeester, D., Marzoli, I., Karman, G. P., Schleich, W. & Woerdman, J. P. Optical Galton board. *Phys. Rev. A* **61**, 013410, (1999).
25   Navarrete-Benlloch, C., Pérez, A. & Roldán, E. Nonlinear optical Galton board. *Phys. Rev. A* **75**, 062333, (2007).
26   Di Molfetta, G., Debbasch, F. & Brachet, M. Nonlinear optical Galton board: Thermalization and continuous limit. *Phys. Rev. E* **92**, 042923, (2015).
27   Gerasimenko, Y., Tarasinski, B. & Beenakker, C. W. J. Attractor-repeller pair of topological zero modes in a nonlinear quantum walk. *Phys. Rev. A* **93**, 022329, (2016).
28   Yao, J. P. Arbitrary waveform generation. *Nat. Photon.* **4**, 79-80, (2010).
29   Cundiff, S. T. & Weiner, A. M. Optical arbitrary waveform generation. *Nat. Photon.* **4**, 760-766, (2010).
30   Yao, J. P. Photonic generation of microwave arbitrary waveforms. *Opt. Commun.* **284**, 3723-3736, (2011).
31   Rashidinejad, A., Li, Y. & Weiner, A. M. Recent Advances in Programmable Photonic-Assisted Ultrabroadband Radio-Frequency Arbitrary Waveform Generation. *IEEE J. Quantum Electron.* **52**, 1-17, (2016).
32   *Keysight M8199A Arbitrary Waveform Generator*, <https://www.keysight.com/us/en/product/M8199A/arbitrary-waveform-generator-128-256-gsas.html>.
33   Ferdous, F. *et al.* Spectral line-by-line pulse shaping of on-chip microresonator frequency combs. *Nat. Photon.* **5**, 770-776, (2011).
34   Weiner, A. M. Femtosecond pulse shaping using spatial light modulators. *Rev. Sci. Instrum.* **71**, 1929-1960, (2000).
35   Veli, M. *et al.* Terahertz pulse shaping using diffractive surfaces. *Nat. Commun.* **12**, 37, (2021).
36   Divitt, S., Zhu, W., Zhang, C., Lezec, J. H. & Agrawal, A. Ultrafast optical pulse shaping using dielectric metasurfaces. *Science* **364**, 890-894, (2019).
37   McKinney, J. D., Leaird, D. E. & Weiner, A. M. Millimeter-wave arbitrary waveform generation with a direct space-to-time pulse shaper. *Opt. Lett.* **27**, 1345-1347, (2002).
38   Yao, J. P. Microwave Photonics. *J. lightw. Technol.* **27**, 314-335, (2009).
39   Marpaung, D., Yao, J. P. & Capmany, J. Integrated microwave photonics. *Nat. Photon.* **13**, 80-90, (2019).
40   Zou, X. *et al.* Photonics for microwave measurements. *Laser Photonics Rev.* **10**, 711-734, (2016).
41   Zhang, J. & Yao, J. P. Time-stretched sampling of a fast microwave waveform based on the repetitive use of a linearly chirped fiber Bragg grating in a dispersive

|     | loop. *Optica* **1**, 64-69, (2014). |
| --- | --- |
| 42  | Meng, J., Miscuglio, M., George, J. K., Babakhani, A. & Sorger, V. J. Electronic Bottleneck Suppression in Next-Generation Networks with Integrated Photonic Digital-to-Analog Converters. *Adv. Photon. Res.* **2**, 2000033, (2020). |
| 43  | Zhang, T., Qiu, Q., Su, J., Fan, Z. & Xu, M. Optical assisted digital-to-analog conversion using dispersion-based wavelength multiplexing. *Opt. Commun.* **432**, 44-48, (2019). |
| 44  | Yacoubian, A. & Das, P. K. Digital-to-analog conversion using electrooptic modulators. *IEEE Photon. Technol. Lett.* **15**, 117-119, (2003). |
| 45  | Peng, Y., Zhang, H., Zhang, Y. & Yao, M. Photonic Digital-to-Analog Converter Based on Summing of Serial Weighted Multiwavelength Pulses. *IEEE Photon. Technol. Lett.* **20**, 2135-2137, (2008). |
| 46  | Gao, B., Zhang, F. & Pan, S. Experimental demonstration of arbitrary waveform generation by a 4-bit photonic digital-to-analog converter. *Opt. Commun.* **383**, 191-196, (2017). |
| 47  | Okada, T., Kobayashi, R., Rui, W., Sagara, M. & Matsuura, M. Photonic digital-to-analog conversion using a blue frequency chirp in a semiconductor optical amplifier. *Opt. Lett.* **45**, 1483-1486, (2020). |
| 48  | Liu, W. *et al.* An integrated parity-time symmetric wavelength-tunable single-mode microring laser. *Nat. Commun.* **8**, 15389, (2017). |
| 49  | Liu, W. *et al.* A fully reconfigurable photonic integrated signal processor. *Nat. Photon.* **10**, 190-195, (2016). |
| 50  | Chi, H., Wang, C. & Yao, J. P. Photonic Generation of Wideband Chirped Microwave Waveforms. *IEEE J. Microwav.* **1**, 787-803, (2021). |
| 51  | Tan, M. *et al.* Photonic RF Arbitrary Waveform Generator Based on a Soliton Crystal Micro-Comb Source. *J. lightw. Technol.* **38**, 6221-6226, (2020). |
| 52  | Xie, Q., Zhang, H. & Shu, C. Programmable Schemes on Temporal Waveform Processing of Optical Pulse Trains. *J. lightw. Technol.* **38**, 339-345, (2020). |
| 53  | Hao, T. *et al.* Breaking the limitation of mode building time in an optoelectronic oscillator. *Nat. Commun.* **9**, 1839, (2018). |
| 54  | Tang, J. *et al.* Hybrid Fourier-domain mode-locked laser for ultra-wideband linearly chirped microwave waveform generation. *Nat. Commun.* **11**, 3814, (2020). |
| 55  | Zhang, J. & Yao, J. P. Parity-time-symmetric optoelectronic oscillator. *Sci. Adv.* **4**, eaar6782, (2018). |

# Supplementary Materials

**Supplementary Material 1: transfer matrix**

Figure S1(a) shows the photonic Galton board (PGB) in which a seed pulse is propagating in the first three round trips, and the generation of a waveform with four sampling points is shown in Figure S1(b). In Figure S1(a), the arrowed blue and red lines indicate the paths by which the pulses travel in the long and short loops, respectively. The beam splitters (BSs) are depicted as gray rectangles. The orange and green circles are the acousto-optic modulators (AOMs) in the long and short loops, respectively. In Figure S1(b), the orange lines show the temporally sampled waveform with four sampling points generated after the 3$^{rd}$ roundtrip.

A 2×2 BS, in free space optics form and fiber optics form shown in Figure S2(a) and (b), respectively, has a transfer matrix given by

$$\begin{pmatrix} b_1 \\ b_2 \end{pmatrix} = \begin{pmatrix} \frac{\sqrt{2}}{2} & i\frac{\sqrt{2}}{2} \\ i\frac{\sqrt{2}}{2} & \frac{\sqrt{2}}{2} \end{pmatrix} \begin{pmatrix} a_1 \\ a_2 \end{pmatrix} \quad (S1)$$

where the imaginary symbol $i$ indicates a $\pi/2$ phase change when the optical pulse from one path propagates into the other path. Thus, the electric field of the optical pulse at the output of the two loops can be expressed by

$$u_{m+1,n} = G_{u,m+1}\left(\frac{\sqrt{2}}{2}u_{m,n-1} + i\frac{\sqrt{2}}{2}v_{m,n}\right) \quad (S2.1)$$

$$v_{m+1,n} = G_{v,m+1}\left(i\frac{\sqrt{2}}{2}u_{m,n-1} + \frac{\sqrt{2}}{2}v_{m,n}\right) \quad (S2.2)$$

where $G_{u,m+1}$ and $G_{v,m+1}$ are the gain/loss of the AOMs in the long and short loops for the ($m$+1)-th round trip, respectively. The output pulse sequence of the two loops at the $m$-th round trip can be expressed in vectors with $m$ elements, given by

$$U_m = [u_{m,1}, u_{m,2}, u_{m,3}, \cdots, u_{m,m}]^T \quad (S3.1)$$

$$V_m = [v_{m,1}, v_{m,2}, v_{m,3}, \cdots, v_{m,m}]^T \quad (S3.2)$$

If we combine $U_m$ and $V_m$, and define another vector $P_m$ with 2$m$ elements, which contains all the pulses in the two loops, we have

$$P_m = [v_{m,1}, u_{m,1}, v_{m,2}, u_{m,2}, v_{m,3}, u_{m,3}, \cdots, v_{m,m}, u_{m,m}]^T \quad (S4)$$

($P_3^T$ is shown as the grey vector at the bottom of Fig. S1(a)). Then, $P_{m+1}$ can be calculated from $P_m$ by a transfer matrix $T_m^{m+1}$ with a scale of 2($m$+1)×2$m$, which can be expressed by

$$T_m^{m+1} = A_m^{m+1} \cdot B_m^{m+1} = \begin{pmatrix} G_{v,m+1} & & & & & & & \\ & G_{u,m+1} & & & & & & \\ & & G_{v,m+1} & & & & & \\ & & & G_{u,m+1} & & & & \\ & & & & \ddots & & & \\ & & & & & G_{v,m+1} & & \\ & & & & & & G_{u,m+1} & \\ & & & & & & & G_{v,m+1} \\ & & & & & & & & G_{u,m+1} \end{pmatrix}_{2(m+1)\times 2(m+1)} \cdot \begin{pmatrix} \sqrt{2}/2 & 0 & 0 & \cdots & & & 0 \\ i\sqrt{2}/2 & 0 & 0 & & & & \\ 0 & i\sqrt{2}/2 & \sqrt{2}/2 & & & & \\ 0 & \sqrt{2}/2 & i\sqrt{2}/2 & & & & \\ \vdots & & & \ddots & & & \vdots \\ & & & & i\sqrt{2}/2 & \sqrt{2}/2 & 0 \\ & & & & \sqrt{2}/2 & i\sqrt{2}/2 & 0 \\ & & & & 0 & 0 & \sqrt{2}/2 \\ 0 & & & \cdots & 0 & 0 & i\sqrt{2}/2 \end{pmatrix}_{2(m+1)\times 2m} \quad (S5)$$

where $A_m^{m+1}$ and $B_m^{m+1}$ are the transfer matrices of the gain/loss of the AOMs and the BS from the $m$-th to the $(m+1)$-th, respectively. For $m = 0$, we have

$$T_0^1 = A_0^1 \cdot B_0^1 = \begin{pmatrix} 1 & 0 \\ 0 & 1 \end{pmatrix} \cdot \begin{pmatrix} i\sqrt{2}/2 \\ \sqrt{2}/2 \end{pmatrix} \quad (S6)$$

Thus, all the pulses in the two loops at the $m$-th round trip can be recursively calculated from the seed pulse. By digitally combining the pulses at the same time slot, as shown in Fig. S1(b), a temporally sampled arbitrary waveform $s[n]$ is generated, as shown at the bottom of the PGB, which can be expressed by

$$s[n] = |u_{m,n}|^2 + |v_{m,n+1}|^2, \quad 0 \leq n \leq m \quad (S7)$$

where $u_{m,0} = 0$ and $v_{m,m+1} = 0$. The sampling interval is equal to the time delay difference between the two loops.

**Fig. S1. Seed pulse in the PGB.** (a) A seed pulse propagating in the photonic Galton board (PGB) in the first three round trips and (b) the generation of a waveform with four sampling points. In (a), the arrowed blue and red lines indicate the pulses traveling in the long and short loops, respectively. The BSs are depicted as gray rectangles. The AOMs in the long and short loops are depicted as orange and green circles, respectively. In (b), a temporally sampled waveform with four sampling points output from the 3$^{rd}$ round trip is shown in the orange lines. The sampling interval is equal to the time delay difference between the two loops

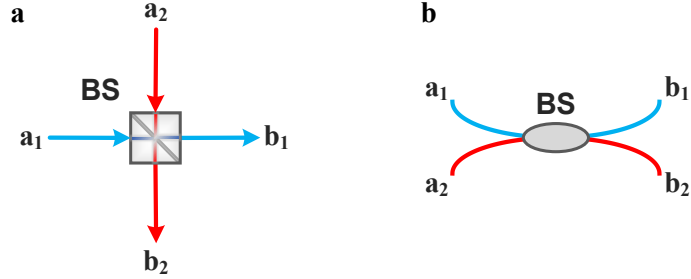

**Fig. S2. Beam splitter.** A 2×2 beam splitter in (a) free space optics form and (b) fiber optics form. $a_1$ and $a_2$ are the two input signals and $b_1$ and $b_2$ are the two output signals.

# Supplementary Material 2: calculating the optimal $G_{u,m}$

Five steps are involved in the calculation of $G_{u,m}$, as shown in Figure S3(a). Firstly, in Step 1, we define a target waveform, $s_t[n]$, with $j$ sampling points and randomly generate 10,000 different $G_u^{(k)}$, each of which is a vector consisting of the $G_{u,m}$ for different round trips. It is shown as

$$G_u^{(k)} = \left[ G_{u,1}^{(k)}, G_{u,2}^{(k)}, G_{u,3}^{(k)}, \cdots, G_{u,j-1}^{(k)} \right]^\mathrm{T}, \quad 1 \leq k \leq 10000 \tag{S8}$$

The distribution of all the $G_{u,m}^{(k)}$ ($1 \leq m \leq j-1, 1 \leq k \leq 10000$) is shown in Figure S4. We use these $G_u^{(k)}$ as the inputs for the numerical model of the PGB to calculate the corresponding waveforms $s[n]^{(k)}$, as shown in Fig. S3(b). These 10,000 pairs of $G_u^{(k)}$ and $s[n]^{(k)}$ form a database. In Step 2, we compare the target waveform with the database. In Step 3, by calculating the root mean square error (RMSE): R of every $s[n]^{(k)}$ in the database with the target waveform, we find a $s[n]^{(c)}$, which has the smallest RMSE (marked it as $R_s$), and the corresponding $G_u^{(c)}$. After that, in Step 4, $G_u^{(c)}$, $R_s$, and $s_t[n]$ are sent into an *Auto-Fit* system with a backpropagation algorithm, with the principle illustrated in Fig. S3(c).

In the *Auto-Fit* operation, we first assign $R_s$ to an initial R, then set $m = 1$ and add a detune $\delta$ to $G_{u,1}^{(c)}$. Next, we send the new $G_u^{(c)}$ to the numerical model of the PGB to get the newly generated waveform and calculate the new RMSE: $R$ between the newly generated and the target waveform. If $R$ is smaller than the previous $R$, adding 1 to $m$ and doing the same operation on $G_{u,m}^{(c)}$ until $m$ is larger than $a$. If not, $R$ becomes larger, taking $2\delta$ from $G_{u,1}^{(c)}$ and sending the new $G_u^{(c)}$ to the PGB again to get a newly generated waveform, then calculate the new RMSE: $R$ between the newly generated and the target waveform. If this new R is smaller than the previous R, add 1 to $m$ and do the same operation on $G_{u,m}^{(c)}$ until $m$ is larger than $a$. If $R$ still becomes larger than the previous, adding $\delta$ to $G_{u,1}^{(c)}$ and refreshing $m$ by $m = m+1$, then doing the same operation on $G_{u,m}^{(c)}$ until $m$ is larger than $a$.

When *m* is larger than *a*, if the present *R* does not equal $R_s$, we set *R* as a new $R_s$ and reset *m* to 1, then repeat the fitting operation from *m* = 1 to *m* = *a*, again. If the present *R* equals $R_s$, we refresh *δ* to be smaller by a coefficient *Ω* for more delicate adjustment of $G_{u,m}^{(c)}$ in the next fitting operation until *δ* is smaller than a predefined threshold. Finally, all the elements in $G_u^{(c)}$ are the optimal $G_{u,m}^{(c)}$ for the generation of the target waveform.

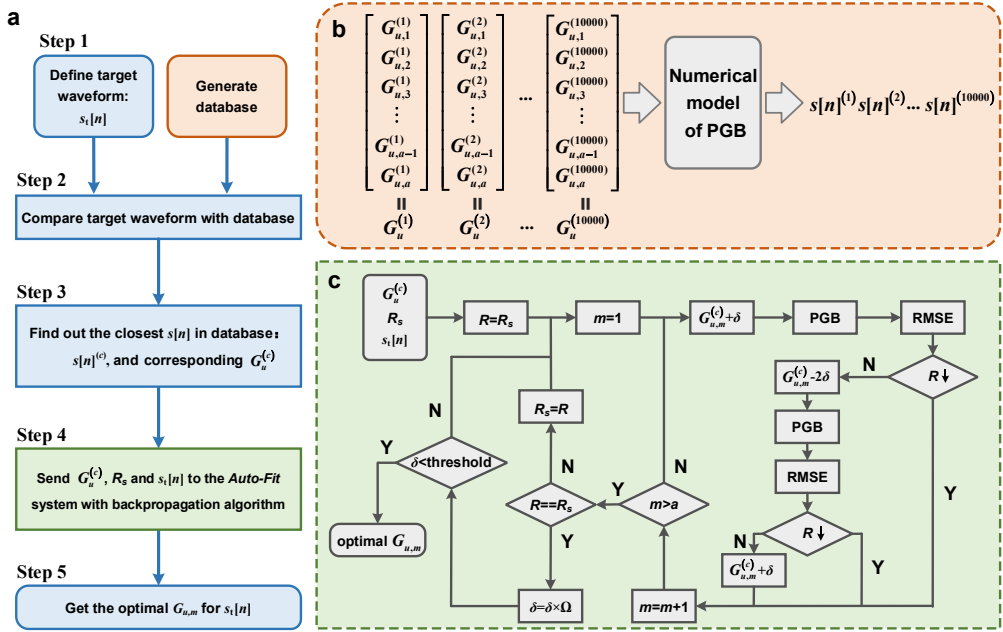

**Fig. S3. The five steps for the calculation of $G_{u,m}$.** (a) The five steps. Start at defining the target waveform and database generating and end up with the optimal $G_{u,m}$. (b) The illustration of database generation. (c) The principle of the *Auto-Fit* system with backpropagation algorithm. Y: yes, N: no, ↓: smaller than previous.

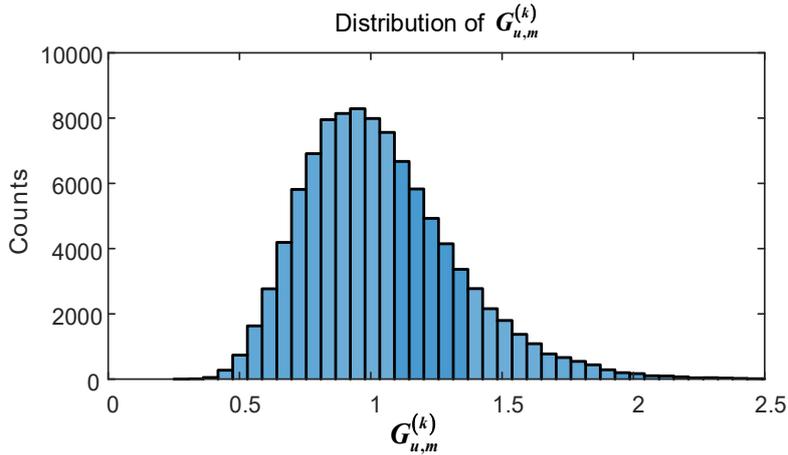

Fig. S4. The distribution of $G_{u,m}^{(k)}$ $(1 \leq m \leq a,\ 1 \leq k \leq 10,000)$.

# Supplementary Material 3: seed pulse, modulation signals and pulse sequences

In the experimental results in Fig. 5 of the Main text, a femtosecond optical pulse from a repetition-rate-reduced pulse train is broadened to have a 3 dB temporal width of 31 ps or a 10 dB width of 56 ps through the SMF with a length of 330 m, as shown in Figure S5(a). For the faster-speed waveform generation shown in Fig. 6 of the Main text, the femtosecond optical pulse after the MZM is injected into the long loop as a seed pulse without broadening, as shown in Fig S5(b), which shows a 3 dB temporal width of 11 ps or a 10 dB width of 19 ps on the OSC.

In our experimental setup, we set a constant loss, $G_{v,m} = 0.5$, in the short loop, by fixing the driving signal applied to AOM2 in the short loop to be a sinusoidal signal at 200 MHz with fixed amplitude. At the same time, a variable loss is applied to the long loop for different round trips by driving the AOM1 in the long loop with a second sinusoidal signal, which has a carrier frequency of 80 MHz. It is amplitude modulated by $G_{u,m}$ at 5.78 MHz, as shown in Figure S6(a)-(h).

Figure S7 shows the pulse sequences in the two loops in the case of generating a burst waveform with 31 sampling points. There is a round-trip time delay of about 173 ns. The insert figures show the pulse sequences at the 30th roundtrip of the long and short loops. A designated time window is applied by the AOM3 in the DU to exclude the seed pulse and the pulse sequences after the 43rd or the 44th round trips.

Noted that a target waveform with $j$ sampling points is generated after the $(j-1)$-th round trip, which establish time $t_e$ is given by

$$t_e = (j-1) \times \frac{l}{v} \quad (S9)$$

where $l$ and $v$ are the length of the long loop and the velocity of light in the fiber, respectively.

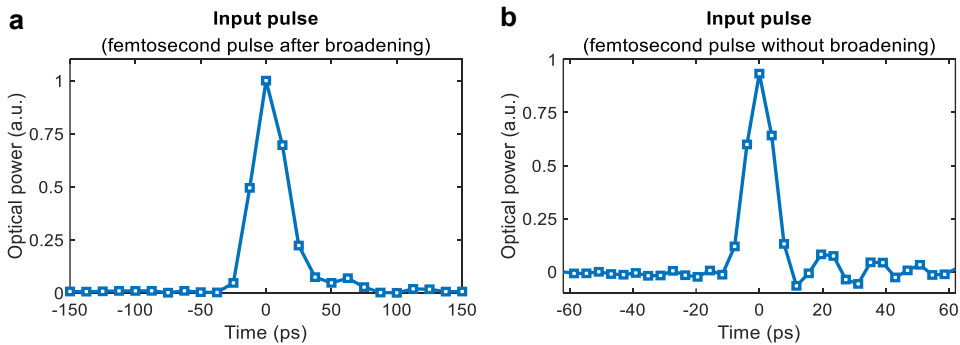

**Fig. S5. Seed pulse.** (a) The femtosecond pulse, after broadening through the SMF with a length of 330 m, has a 3 dB temporal width of 31 ps or a 10 dB width of 56 ps, which is captured on an OSC with a sampling rate of 80 GSa/s. (b) The femtosecond pulse generated from the MLL without broadening is shown on another OSC, which has a higher sampling rate of 256 GSa/s. Each hollow square is a sampling point of the OSC.

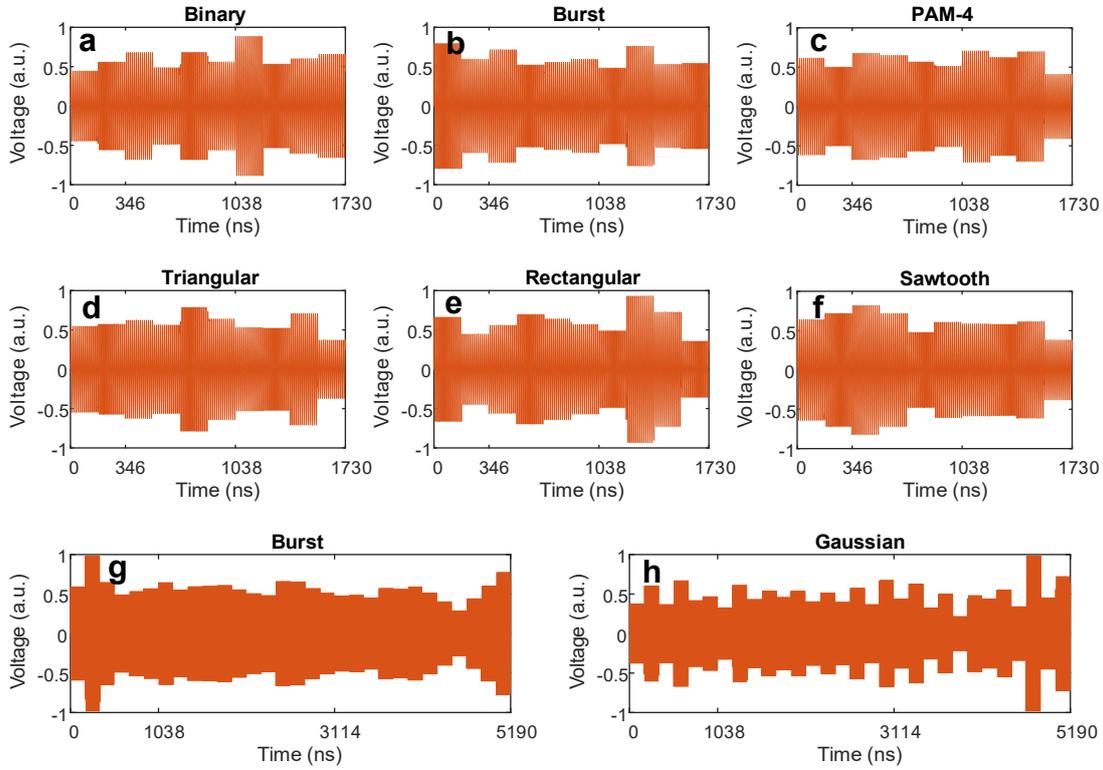

**Fig. S6. Normalized driving signal applied to AOM1 in the long loop.** Sinusoidal signals amplitude modulated by $G_{u,m}$ at 5.78 MHz for the generated waveforms: (a) binary coded, (b) burst, (c) PAM-4, (d) triangular, (e) rectangular, (f) sawtooth, (g) burst and (h) Gaussian. Different amplitudes correspond to the different $G_{u,m}$ for every round trip.

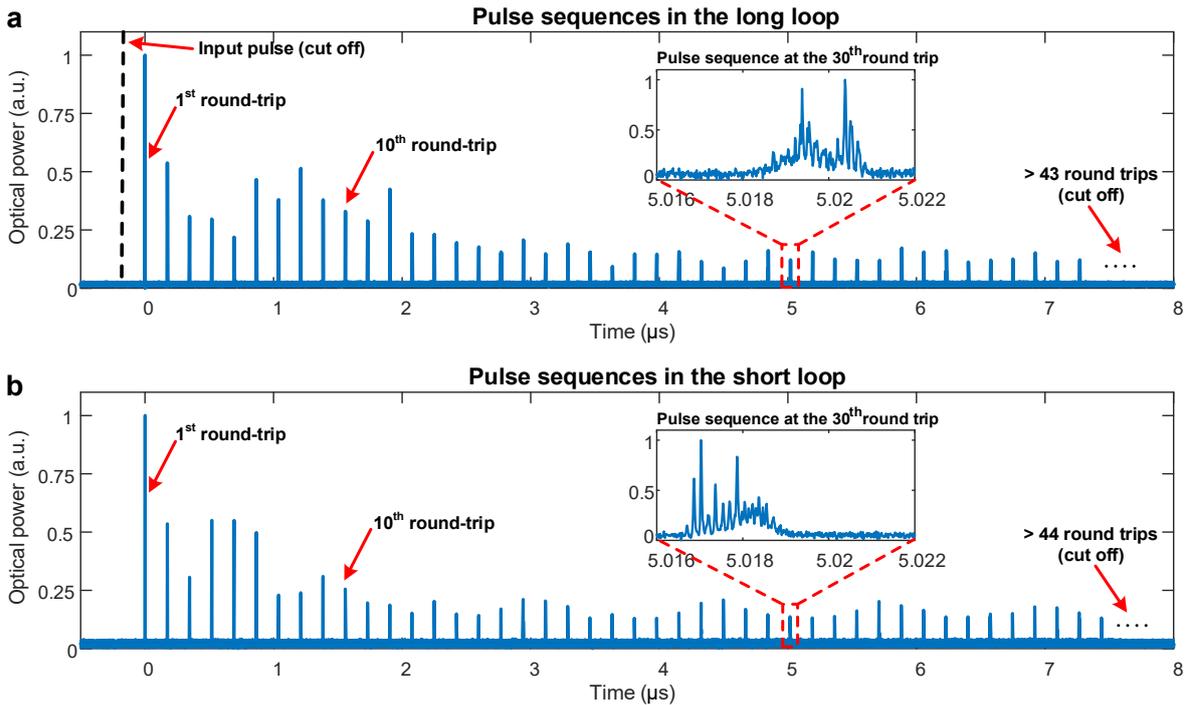

Fig. S7 Pulse sequences from every round trip in the long and the short loops for generating the burst waveform with 31 sampling points. (a) Pulse sequences in the long loop. (b) Pulse sequences in the short loop. The inserts show the pulse sequences at the 30[th] roundtrip in the long and short loops.

## Supplementary Material 4: performance evaluation

We calculate the RMSE of the generated waveforms to evaluate the performance of the PGB for the generation of arbitrary waveforms. The RMSE is given by

$$\text{RMSE} = \sqrt{\frac{\sum_{n=1}^{j}(s_t[n]-s[n])^2}{j}} \quad (S10)$$

where $s_t[n]$ and $s[n]$ are the target waveform and the generated waveform, respectively. $j$ is the number of sampling points in the target waveform.

Table 1 shows the RMSEs of the generated waveforms in Fig. 5 of the Main text, which shows an average value of 0.1820, confirming good fidelity of the generated waveforms. Note that, the RMSEs in Tab. 1 are calculated by extracting the peak values of the sampled pulses in the generated waveforms, and comparing them with the target values of the sampling points in the target waveforms. There are two error symbols in the 31-sampling-point waveforms, which are the largest RMSE among those waveforms. This is caused by the polarization-induced difference in the modulation efficiency of the AOMs and the phase detuning of the optical pulses recirculating in the system.

Table 1. The RMSE of the generated waveforms in Fig. 5 of the Main text

| Waveform | RMSE |
|---|---|
| (a) Binary coded | 0.2425 |
| (b) 11-sampling-points burst | 0.2477 |
| (c) PAM-4 | 0.1170 |
| (d) Triangular | 0.1228 |
| (e) Rectangular | 0.1524 |
| (f) Sawtooth | 0.1197 |
| (g) 31-sampling-points burst | 0.3793 |
| (h) Gaussian | 0.0742 |

As for the faster-speed waveforms with tunable sampling rates from 21.7 to 80.0 GSa/s in Fig. 6 of the Main text, we put the target waveforms into a linear-time-invariant system, which has a passband of 50 GHz and a sampling rate of 256 GSa/s, to emulate the PD and OSC in our experimental setup. Then we use those processed target waveforms to evaluate the performance of the generated waveforms. The corresponding RMSEs are shown in Table 2, which shows an average value of 0.0798, again confirming the good fidelity of the generated waveforms.

Table 2. The RMSE of the generated waveforms in Fig. 6 of the Main text

| Waveform – Sampling rate | RMSE |
|---|---|
| (a) Triangular - 21.7 GSa/s | 0.0691 |
| (b) Rectangular - 48.0 GSa/s | 0.1071 |
| (c) Sawtooth - 31.0 GSa/s | 0.0819 |
| (d) Triangular - 44.1 GSa/s | 0.0999 |
| (e) Rectangular - 64.0 GSa/s | 0.1008 |
| (f) Sawtooth - 51.2 GSa/s | 0.0830 |
| (g) Gaussian - 52.2 GSa/s | 0.0534 |
| (h) Gaussian - 80.0 GSa/s | 0.0427 |

The RMSEs of the rectangular waveforms are a little bit larger than those of the other waveforms, which, to some extent, are caused by the speed of rise/fall slopes of our OSC. In addition, the deviation of all generated waveforms is also caused by the noise figure of our EDFAs in the two loops, the polarization-induced difference of the modulation efficiency during every round trip, the optical nonlinear effect, phase detuning of the optical pulses recirculating in the system and etc.